\documentclass[5p]{elsarticle}

\usepackage[T1]{fontenc}
\usepackage[shortcuts]{extdash}
\usepackage{color}
\usepackage[dvipsnames]{xcolor}
\usepackage{amsmath}
\usepackage{amssymb}
\usepackage{amsfonts}
\usepackage{longtable}
\usepackage{hyperref}
\usepackage{booktabs}
\usepackage{array}
\usepackage{multicol}
\usepackage{graphicx}
\usepackage{ulem}
\usepackage[final]{changes}

\definechangesauthor[color=black]{k}
\definechangesauthor[color=blue]{b}

\graphicspath{{fig/}{NEON/fig/}}

\journal{Astroparticle Physics}

\DeclareMathOperator{\km}{km}

\DeclareMathOperator{\TeV}{TeV}

\DeclareMathOperator{\PeV}{PeV}

\newcommand{\remove}[1]{}

\definechangesauthor[color=blue]{ly}
\definechangesauthor[color=Rhodamine]{zhm}
\definechangesauthor[color=ForestGreen]{lsj}
\definechangesauthor[color=purple]{qiuzj}
\definechangesauthor[color=Goldenrod]{hyl}
\definechangesauthor[color=brown]{tsdisee}
\definechangesauthor[color=orange]{lyh}

\biboptions{sort&compress}

\begin{document}

\begin{frontmatter}

  \title{A proposed deep sea Neutrino Observatory in the Nanhai}

  \author[cga]{Huiming Zhang}
  \author[cga]{Yudong Cui}
  \author[cga]{Yunlei Huang}
  \author[cga]{Sujie Lin\corref{cor1}}
  \ead{linsj6@mail.sysu.edu.cn}
  \author[cga]{Yihan Liu}
  \author[cga]{Zijian Qiu}
  \author[cga]{Chengyu Shao}
  \author[cga]{Yihan Shi}
  \author[cga]{Caijin Xie}
  \author[cga]{Lili Yang\corref{cor1}}
  \cortext[cor1]{Corresponding author}
  \ead{yanglli5@mail.sysu.edu.cn}
  \affiliation[cga]{
    addressline={School of Physics and Astronomy, Sun Yat-sen University},
    city={Zhuhai},
    state={Guangdong},
    postcode={519082},
    country={China}
  }
\hyphenation{Che-ren-kov}
\hyphenation{pro-ject}
\hyphenation{sear-ches}
\hyphenation{recor-ded}

\begin{abstract}%
Over the past ten years, several breakthroughs have been made in multi-messenger astronomy. Thanks to the IceCube Neutrino Observatory, the detection of astrophysical neutrinos was proved to be practical. \replaced{However, due to the limited statistics and field of view, only a few sources have been associated with IceCube neutrinos, making new and larger neutrino telescopes necessary. }{However, no source has been significantly identified due to the lack of statistics and uncovered field of view. The next generation of high-energy neutrino telescopes is in high demand.} We propose the NEutrino Observatory in the Nanhai (NEON), located in the South China Sea to be complementary for the global neutrino detectors.
This proposal describes the design and layout of the array and reports on comprehensive simulations conducted to assess its performance.
The NEON project, with a volume of 10 km$^3$, achieves an angular resolution of 0.1$^\circ$ at 100 TeV.
With 10 years of operation, the project's 5$\sigma$ sensitivity is estimated as $E^2\Phi \sim 3 \times 10^{-10}$ GeV cm$^{-2}$ s$^{-1}$ for a source spectrum index of -2.
We found that the variation in depth from 1700 to 3500 meters does not significantly influence the sensitivity to steady sources.

\end{abstract}


\end{frontmatter}

\settruncatewidth{.25\textwidth}

\section{Introduction}\label{sec:introduction}
Since the birth of neutrino on the 4th December 1930, numerous efforts have been made to explore their properties and origin. The presence of neutrinos is expected to be shown by the particles they produce upon interaction. However, as neutrinos are neutral and hardly interact with matter, they are for all intents and purposes invisible. To catch these invisible particles, the first massive detector was built by Cowan and Reines \cite{Cowan:1956rrn}, marking the beginning of the \replaced{neutrino }{large-scale} facility era.

Until the first collection of 24 neutrinos with a few tens of MeV energies from SN1987A on 23 February 1987, the dawn of neutrino astronomy had come \cite{Kamiokande-II:1987idp, IMB:1987klg}. In the following decades, tons of studies have been performed based on this unique event, not only testing the theories of stellar collapse but also providing knowledge on fundamental physics. Around the same time, space-born detectors observed many gamma-ray sources outside our solar system. Ground-based experiments, \replaced{such as the Pierre Auger Observatory, the Large High Altitude Air Shower Observatory (LHAASO), and the Telescope Array (TA),}{such as the Pierre Auger Observatory and the Large High Altitude Air Shower Observatory (LHAASO),} have surprisingly brought cosmic-ray and gamma-ray detections up to EeV and PeV energies, respectively
\replaced{\cite{PierreAuger:2017pzq, LHAASO:2021cbz, ogio2019telescope}}{\cite{PierreAuger:2017pzq, LHAASO:2021cbz}}.

In the last decade, IceCube neutrino observatory with a volume of $\sim1\km^3$ has shown the existence of extraterrestrial neutrinos \cite{IceCube:2013cdw, IceCube:2013low}, which was later \replaced{tested }{confirmed} by ANTARES \cite{Fusco:2021axs} and Baikal-GVD \cite{Baikal-GVD:2021civ}. These experiments have demonstrated the capability to detect high-energy events within a substantial volume of medium. IceCube has been continuously operating for 13 years and  observed that the spectrum of celestial neutrinos generally \replaced{is consistent with a single }{follows a simple} power law with several selection criteria, \deleted{evenly distributed between flavors, } as seen in Figure~\ref{fig:full_sky_spec}. The measurements with various analyses, such as high energy starting events \cite{IceCube:HESE7.5}, through-going muon tracks \cite{IceCube:2021track}, cascades \cite{aartsen2020characteristics} and enhanced starting track event selection \cite{IceCube:2024estes} are shown. These independent measurements are compatible with each other with spectral indices $\sim$ 2.4 - 2.9. The sources of TXS 0506+056 \cite{IceCube:2018dTXS, IceCube:2018TXSnu}, NGC 1068 \cite{IceCube:2022NGC} and Galactic plane \cite{IceCube:diffuse} have been observed with certain statistical significance . As current evidence shows,  galactic diffuse neutrinos only contribute less than 10 percent to the total observations \cite{Fang:2023azx, Shao:2023aoi}. Most of the astrophysical neutrinos are from extragalactic space. However, no clear associated source was found with one particular catalog. 

Neutrino astronomy now faces an evident lack of statistics and the ability to identify single sources. To make further progress, making a 5 $\sigma$ detection with high statistics, the deployment of a more massive detector is on demand. Some next-generation neutrino telescopes are being constructed and planned, such as IceCube-Gen2 \cite{aartsen2019neutrino}, KM3NeT \cite{adrian2016letter}, \replaced{P-ONE }{P-One} \cite{agostini2020pacific}, TRIDENT \cite{TRIDENT:proposal} and HUNT \cite{HUNT:2023mzt}.
Based on various purposes and focuses, the design, layout and techniques of the telescope are different. A comparison of the design and layout of these projects can be found in Table \ref{tab:exp_comparison}. 

Here we propose the construction of a deep-sea neutrino telescope, the NEutrino Observatory at the Nanhai (NEON), with a volume of around 10 $\km^3$, and the choice of depth of 1700 and 3500 meters in the South China Sea.
\begin{table*}[!htpb]
\centering
\small
\begin{tabular}{lcccc}
\toprule
\textbf{Detector} & \textbf{Volume $[km^3]$} & \textbf{Number of Strings} & \textbf{Number of OMs} & \textbf{Type of OM} \\
\midrule
IceCube & 1 & 86 & 5000 & \textbf{1} $\times$ \textit{10"} PMT \\
Baikal GVD & $\sim$1.5 & 216 & 10,386 & \textbf{1} $\times$ \textit{10"} PMT \\
KM3NeT ARCA & $\sim$1 & 230 & 4140  & \textbf{31} $\times$ \textit{3"} PMTs \\
P-ONE & \>1 & 70 & 1400 & \textbf{16} $\times$ \textit{3.1"} PMTs \\
IceCube Gen2 & 8 & 120 & 9600 & \textbf{16} / \textbf{18} $\times$ \textit{4"} PMTs \\
TRIDENT & 7.5 & 1,211 & 24,000 & \textbf{31} $\times$ \textit{3"} PMTs \& SiPM \\
HUNT & 30 & 2,304 & 55,000 & \textbf{1} $\times$ \textit{20"} PMT \\
NEON & 10 & 1,200 & 21,600 & \textbf{31} $\times$ \textit{3"} PMTs \\
\bottomrule
\end{tabular}
\caption{The physical parameters of operating, constructing and planed neutrino telescopes.}
\label{tab:exp_comparison}
\end{table*}
The NEON project is dedicated to the discovery of neutrino source and the study of neutrino physics, expanding the effective area of the current neutrino telescope by one order of magnitude while preserving the capability to detect TeV neutrino events. Therefore, NEON will largely increase the event rate with a wide energy coverage, large sky coverage, and improved angular resolution. This would enable the confirmation of neutrino candidate sources and the potential discovery of transient and steady sources, especially in the southern hemisphere where IceCube observation \replaced{is restricted to higher energies }{fails to cover}. NEON has the potential to \replaced{provide further validation of the fundamental physics }{address fundamental physics questions}, such as \deleted{the detection of} the Glashow resonance, and to provide insights into Earth and marine sciences, including Earth tomography.

In this proposal, we first state the scientific targets of the NEON project in Section \ref{science}. These include but are not limited to the search for extragalactic and galactic neutrino sources. We present the design of the NEON project in Section \ref{sec:telescope}, where the detection principle, the array layout, the optical noise in the deep sea, and the functional optical module are described. Section \ref{sec:simulation} introduces the comprehensive simulation framework, NEONSim, which comprises background simulation, neutrino interaction simulation, Cherenkov radiation simulation, and optical module detection simulation. The performance of NEON, including its angular resolution, effective area, and sensitivity, is presented in Section \ref{performance}. Finally, discussions are given in Section \ref{sec:summary}.

\section{Scientific targets}\label{science}
\begin{figure}
    \centering
    \includegraphics[width=0.45\textwidth]{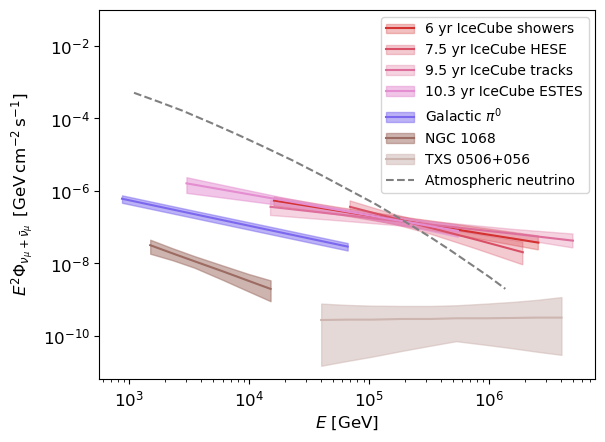}
    \caption{
    The figure shows fluxes of high-energy muon neutrino above 1 TeV from various sources.
    The series of different red bands represent IceCube's all-sky observations with various selection criteria \cite{IceCube:2020acn,IceCube:2021track,IceCube:2021uhz,IceCube:2024estes}.
    The blue band shows the best-fitting result of the $\pi^0$ model for Galactic plane \cite{IceCube:diffuse}.
    The brown and light brown bands are the observed spectrum of two point sources \cite{IceCube:2022NGC}.
    The background atmospheric neutrino is shown as gray dashed line  \cite{IceCube:2014slq}.
    }
    
    \label{fig:full_sky_spec}
\end{figure}

The first observation of astrophysical neutrinos in the TeV to PeV range in 2013 by IceCube marks a milestone of high-energy neutrino astronomy~\cite{IceCube:2013cdw}. \added{However, despite significant efforts, identifying specific sources of these neutrinos has proven challenging.} Until now,\replaced{ even though studies on the association between neutrino events and point sources, as well as the galactic plane, have been published, more observations are needed to achieve statistically significant results }{no statistically significant evidence of directional clustering has been found for the point-source search}. The isotropic distribution of neutrino events with a hard energy spectrum of $E_{\nu}^{-\gamma}$ indicates an extragalactic dominant origin.
The first observation of astrophysical neutrinos in the TeV to PeV range by IceCube in 2013 marked a milestone in high-energy neutrino astronomy~\cite{IceCube:2013cdw}.  Until now, no statistically significant evidence of directional clustering has been found for point-source search.
Some efforts have been made to locate and identify the sources. For instance, stacking analysis for correlation between neutrino events and known gamma-ray catalogs, and point source analysis for both time-dependent transient and steady sources have been performed \cite{IceCube:2023htm, IceCube:2023atb, IceCube:2023oqe}. 
The expected atmospheric neutrino flux and the measurements of total diffuse neutrino fluxes with various selection criteria, and two sources of TXS 0506+056 and NGC 1068 are shown in Figure~\ref{fig:full_sky_spec}. At energies above 100 TeV, the astrophysical neutrinos are dominant. These two AGNs are with different spectral features and each only contribute less than 1\% to the total neutrino flux. This indicates there are more unidentified sources in the Universe. 

Recent breakthroughs from the LHAASO project in China have yielded the observation of the emissions from tens of sources within the Milky Way above 25 TeV, which also extend the gamma spectral energy distribution (SED) up to PeV high energies \cite{LHAASO:2023rpg}. 
Current multi‐wavelength astronomy from radio waves to gamma rays has given more potential neutrino sources with non‐thermal radiation, from extragalaxy, like blazars, gamma‐ray bursts, etc; and from our Galaxy such as supernova remnants and microquasars. Moreover, \replaced{the observations of deep-sea neutrino telescopes }{neutrino observation} can provide hints on fundamental physics as dark matter and neutrino physics and investigate the fields of marine and Earth sciences.

\subsection{Neutrino sources} \label{subsec:neutrino_sources}

The TeV - PeV astrophysical neutrino diffuse flux could be explained by the contribution from our Galaxy and outside our galaxy. \deleted{The physical scenarios for producing these neutrinos are proton-proton and proton-photon interactions. Neutrinos are generated when high-energy cosmic rays interact with gas or photons on site or during the propagation.} \added{There is a natural connection between cosmic rays, gamma rays, and neutrinos. When extreme astrophysical phenomena occur, some of the gravitational energy released in the accretion of matter is transformed into the acceleration of protons and heavy nuclei. These charged particles will interact with ambient radiation, matter, or molecular clouds to generate pions and other secondary particles which will decay into neutrinos, such as $p + \gamma \to \pi^0 + p $ and $p + \gamma \to \pi^{+} + n $. Neutral pions decay into two gamma rays, $\pi^0 \to \gamma + \gamma$. The charged pions produce three neutrinos and one positron from the pion decay chain. The correlation searches between neutrinos and many known catalogs have been performed for several decades. Even though no evident association has been found, no doubtfully the era of multi-messenger astronomy has arrived with the construction and operation of larger and more sensitive facilities across the electromagnetic spectrum reinforces the multi-wavelength (MW) approach in the field.}

A. Extragalactic origin 

The isotropic distribution of the neutrino arrival directions favors an extragalactic origin. Moreover, \replaced{any deviation from a single power law may indicates that some different types of distant sources may contribute }{the general simple power-law spectrum indicates that some different types of distant sources may contribute}. Based on the current knowledge, especially from electromagnetic observations, correlation studies between neutrino events and known sources and stacking analysis to the same catalog have been well established.

The most popular catalogs are from Fermi-LAT, such as 4FGL and 4LAC \citep{Fermi-LAT:2019yla, Fermi-LAT:2019pir}, where Active Galactic Nuclei (AGN) dominate. Blazar as a subclass of AGN whose jets point toward Earth takes a lot attention since the coincident in time and direction of IceCube-170922A  with the gamma-ray \added{flare of} blazar TXS 0506+056 was found \citep{IceCube:2018TXSnu}. Later, IceCube observed an excess in their source list search at the location of the Seyfert II galaxy NGC 1068 at the 4.2$\sigma$ level with cumulative analysis \citep{IceCube:2022NGC}. Gamma-ray bursts (GRBs) are long-believed potential neutrino sources. They release a total energy of about $10^{44}$ J within a few tenths of a second to hundreds of seconds, making it the most violent explosion in the universe. High-energy neutrinos can be produced in both prompt and afterglow phases, spanning from MeV to PeV energy range. Stringent constraints have been placed \citep{IceCube:2023rhf}, which do not favor the origin of GRB. Tidal disruption event (TDE), occurring when a star passes close to a massive black hole and is destroyed by its gravity, is another extreme phenomenon. In their relativistic jets, neutrinos as counterparts will be generated and escape from the site. The recent study found an association of a track-like astrophysical neutrino IceCube 191001A with a tidal disruption event AT2019dsg showing the possibility of TDE as the neutrino source\citep{Stein:2020xhk, Winter:2020ptf}.

The statistical significance for normal correlation searches is weakened due to the uncertainty of the time window and searching region. The efficient method of identifying neutrino sources is the transient point source analysis, where the background will be dramatically reduced because of the short duration of these events. \replaced{In the typical time window of transient source, the expected number background events is less than $10^{-2}$. According to the Feldman-Cousins approach\cite{feldman1998unified}, if }{If} two or more neutrino candidates (doublet and triplet) from the same direction within a certain time are observed, the confidence level will be high enough to confirm the observation.

As ultra-high-energy cosmic rays (UHECRs) with energies of EeV and above propagate from distant sources, they will interact with cosmic background photon fields, such as cosmic microwave background (CMB) and extragalactic background light (EBL). The counterpart neutrinos in the energy range of PeV to EeV are produced along the path via photopion interaction. Since the Pierre Auger Observatory has been successfully observing these UHECRs for almost two decades, these cosmogenic neutrinos are guaranteed sources. \\

B. Galactic sources

The current ground-based gamma-ray experiments have achieved big success, with better precision and extending the SED to the PeV energy region. However, no PeVatron has been identified, because gamma rays can be produced via both leptonic and hadronic processes. Neutrino as a smoking gun will play a key role in locating the origin of Galactic cosmic rays. 

For a long time, supernova remnants have been considered as one of the most possible candidates for the source of galactic cosmic rays.
High-energy cosmic rays could collide with the surrounding gas and produce gamma rays and neutrinos above TeV if molecular clouds are present within 100 pc of a supernova. 
The observations of IC 443, W 44, and W 51 have indeed shown the rationality of the hadronic model \citep{Fermi-LAT:2013iui, MAGIC:2012anb}.

Although it is widely accepted that pulsars and pulsar wind nebulae are of leptonic origin, some theories suggest the presence of relativistic hadrons in the pulsar wind.
The magnetic field of pulsars is able to trap cosmic rays up to 3 PeV, and thus producing high-energy gamma rays and neutrinos.
For more detailed studies, see references \cite{Guetta:2002hv, Amato:2003kw, Amato:2006ts, Bednarek:2003cv, Lemoine:2014ala}.

Another popular Galactic neutrino candidate is young massive star clusters which are located in molecular clouds and formed by the aggregation of young stars.
Cosmic rays emitted by supernova remnants born within these clusters \citep{Ambrosone:2023hsz, Bykov:2020zqf} may readily collide with the surrounding stellar material, producing high-energy gamma rays and neutrinos.

When the gas environment around a hadronic source becomes dense enough to capture the emitted gamma rays, the source may only release neutrinos, potentially accompanied by X-rays or radios.
These sources are referred to as hidden sources.
With the accumulation of experimental statistics, they can yield a detectable signal in the stacking analysis of neutrino sources.

\subsection{Multi-messenger astronomy} \label{subsec:mma}
On 22 September 2017, IceCube detected a neutrino event with 290 TeV energy, and an alert was sent to the Gamma-Ray Burst Coordinates Network (GCN) and the Transient Astronomy Network (TAN). Within the angular error of this neutrino event, several blazars were identified, among which TXS 0506+056 was identified by Fermi-LAT as being in a flaring state and was simultaneously observed by the MAGIC telescope. Afterward, about 20 astronomical devices around the world conducted follow-up observations around that position and found electromagnetic signals in other bands \cite{IceCube:2018dnn}. Excitingly, the detection of two messengers from the same source, high-energy neutrinos and gamma rays indicates that blazars are possible sources of high-energy neutrinos.

\deleted{There is a natural connection between cosmic rays, gamma rays, and neutrinos. When extreme astrophysical phenomena occur, some of the gravitational energy released in the accretion of matter is transformed into the acceleration of protons and heavy nuclei. These charged particles will interact with ambient radiation, matter, or molecular clouds to generate pions and other secondary particles which will decay into neutrinos, such as $p + \gamma \to \pi^0 + p $ and $p + \gamma \to \pi^{+} + n $. Neutral pions decay into two gamma rays, $\pi^0 \to \gamma + \gamma$. The charged pions produce three neutrinos and one positron from the pion decay chain. The correlation searches between neutrinos and many known catalogs have been performed for several decades. Even though no evident association has been found, no doubtfully the era of multi-messenger astronomy has arrived with the construction and operation of larger and more sensitive facilities across the electromagnetic spectrum reinforces the multi-wavelength (MW) approach in the field.}

\added{As electromagnetic emission is the most conventional probe for astrophysical objects. The success of electromagnetic counterparts observation to both high-energy neutrinos and gravitational waves, as IC-170922A and GW 170817 \cite{IceCube:2018dTXS, LIGOScientific:2017vwq}, demonstrated the power of multi-messenger approach in astrophysics. It led to synergies and cooperation of a global network, which is essential to integrate and optimize the MW and MM instruments. In particular, with the major breakthrough in the field, LHAASO greatly complemented the absence of high-energy parts of the SED. }

\added{Moreover, as the first report of gravitational wave (GW) from a stellar mass binary black hole merger GW150914 was announced by laser interferometric gravitational-wave detectors (LIGO) in 2016 \cite{LIGOScientific:2016aoc}, the correlation study between GW and neutrinos has been performed \cite{IceCube:2020xks, IceCube:2022mma}. For compact object with dense environment, both neutrinos and GWs can be emitted, such as neutron-star mergers. Neutrinos can locate a source with an angular resolution smaller than 1 degree, and, more importantly, can probe the source engine even without electromagnetic signal.}

\added{NEON will play an essential role in the era of multi-messenger astronomy by providing neutrinos with more statistic and better positioning. Through synergy with global instruments and combining the information from other messengers, it will be possible to resolve the questions in astrophysics and fundamental physics. }

\section{The design of NEON project} \label{sec:telescope}
Moisey Markov proposed the first generation of an undersea neutrino telescope in 1960 \cite{Markov:1961tyz}.
The idea is to detect neutrinos by measuring the Cherenkov radiation emitted by secondary particles that are created when neutrinos interact with matter. 
Subsequent efforts focused on high-energy neutrino detection were performed, including the Deep Underwater Muon And Neutrino Detector (DUMAND) Project \cite{DUMAND:1988pgt}.

However due to technical difficulties, they were declined unwillingly and DUMAND is marked as the precursor of the second-generation deep-sea neutrino telescope projects, such as AMANDA \cite{Andres:2001ty}, ANTARES \cite{ANTARES:2006cjl},\added{ Baikal-GVD \cite{belolaptikov2021neutrino},} NEMO \cite{Piattelli:2004lxr} and the NESTOR Project \cite{NESTOR:2005jdw}. 

This century, the IceCube experiment with a volume of about $1\km^3$ instrumented at the South Pole has achieved a significant milestone in the observation of astrophysical neutrinos with thirteen years of operation.
The KM3NeT in the Mediterranean Sea and the GVD in the Baikal Lake are in their constructing phase, and both plan to cover 1  km$^3$ volume \cite{belolaptikov2021neutrino} .
The current observations provide limits on the well-known objects and serve as a baseline for future observations.
The future progress requires a next-generation experiment with a sensitivity superior by one order of magnitude.
Aiming at revealing the open questions of both astronomy and physics, and reaching the requirement of detection sensitivity, the NEON project with two optional sites at 1700 and 3500 meters is designed with larger volume and better resolution to fulfill the demand.

Neutrinos with three flavors interact with the medium via charged current (CC) or neutral current (NC) interactions and produce secondaries emitting Cherenkov light.
This Cherenkov cone is aligned to the direction of the particle motion with an opening angle that depending on the particle velocity and the refractive index of the medium $\cos\theta_{\mathrm{Che}} = \frac{c}{nv}$.
These light cones are collected by photosensitive devices in a large transparent medium.  

To obtain the energy and direction of neutrinos with high statistics, a 3-dimension optical module (OM) array in gigantic transparent seawater is planned.
The layout of the OMs within the array is designed with the impacts of light absorption and scattering in the sea water.

In order to eliminate the noise hits arising from bioluminescence and potassium-40 ($^{40}\mathrm{K}$),  each OM will incorporate tens of PMTs to capture more directional information of the light.
Both onshore and offshore tests have been conducted on the OM design at the current stage.

\subsection{Detection principle}

Neutrinos only  interact through weak force, and their interaction cross-section with ordinary matter is extremely small.
This fact necessitates the use of expansive targets in the detection of neutrinos to accumulate a sufficient number of collision signals.
One of the most efficient methods to detect the secondaries produced within such extensive targets is to measure the Cherenkov light emitted by these particles.
Consequently, neutrino observatories utilize vast quantities of transparent medium, such as water or ice, as the detector's target, and install photomultiplier tubes (PMTs) within this medium to capture the Cherenkov radiation.

When a neutrino interacts with the nucleon in matter, it can \replaced{trigger }{transfer momentum to the nucleon, triggering} a hadronic cascade.
In the NC interaction, the original neutrino loses momentum and escapes without leaving any other visible signature, $\nu_{\mu(e,\tau)} + N \to \nu_{\mu(e,\tau)}$ + hadrons. The hadrons gained the energy will produce cascade showers. 
By contrast, in one CC interaction, the primary neutrino is converted into a corresponding lepton, generating distinct signatures that depend on the lepton flavor, $\nu_{\mu (e,\tau)} + N \to \mu (e,\tau) + \mathrm{hadrons}$.
At high energies, the resulting lepton retains a direction approximately parallel to the primary neutrino.
Electron, with its short mean-free path in matter, produces an electromagnetic cascade at the interaction vertex.
Tau, which decays close to the vertex, produces an extra cascade.
Muon, being more stable and having a longer mean-free path in matter, creates a unique track-like signature whose direction can be precisely reconstructed, providing better directional information of neutrino.
Therefore, to identify the astrophysical neutrino origin, muon neutrinos are the best probe to track back. The analysis of muon neutrino track-like signatures, which offer the greatest pointing accuracy to the sources is discussed in this work.

The muonic track-like signature is relatively easy to detect, as the muon's extended path passes nearby multiple OMs within the array. 
In contrast, taus decay at a distance from their initial interaction point, showing a ``double bang" morphology.
This decay interval is determined by the $\Gamma$ factor of the secondary tau, which approximates 50 meters for an energy of 1 PeV.
Shower events, however, pose a greater challenge.
They arise from various sources, including secondary electrons from CC interactions and hadronic secondaries from NC interactions.

\subsection{Detector layout} \label{subsec:detector_layout}

It is anticipated that future detectors, with volumes surpassing several $\km^3$ and better sensitivity, will reveal more neutrino sources, thereby heralding a new era in the study of neutrinos.
Consequently, we propose a neutrino detector array containing a physical volume of 10 $\km^3$, with a radius of 1.79 km.

In order to effectively collect sufficient Cherenkov photons for track- or shower-like events, the distance between OMs inside the array need to be determined accordingly to the optical properties of the water.
The attenuation length of light with wavelengths between 400 and 500 nm in the deep sea is approximately 20 to 30 m, as reported by previous studies of TRIDENT \cite{TRIDENT:proposal}. The NEON team has shipped to the site in 2022 and measured the Rayleigh scattering length with the ECO 3-Measurement Sensor\cite{commtec_2007}.
At the depth of 2000 m, the Rayleigh scattering length for the wavelength of 532 nm is 28.8 m, which matches other measurements.
Considering this attenuation length and event size, a horizontal OM spacing of approximately 100 meters ensures that the nearest OM captures at least $5\%$ of the photons, and a vertical OM spacing of about 40 meters is sufficient for maintaining the quality of reconstruction. We refer to the paper \cite{Zhang:2023icn} for more details. 
In this proposal, we thus select a vertical spacing of 40 meters for OMs, which are arranged within strings that extend up to 1 kilometer in length.
Across each string, there are a total of 18 OMs.
Specifically, the array will be composed of 1200 strings and house a total of 669,600 PMTs.

\begin{figure}[!htb]
    \includegraphics[width=0.45\textwidth]{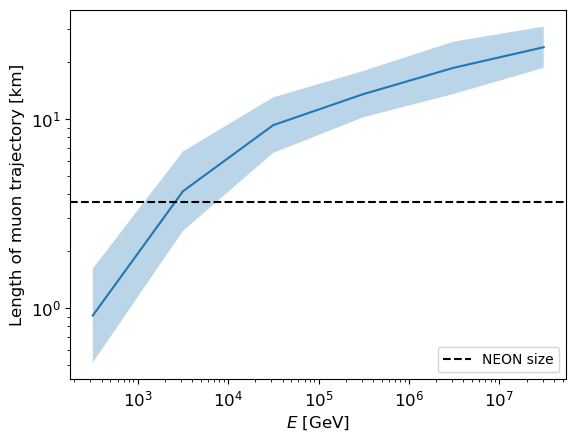}
    \caption{The simulated trajectory length of muons propagating through water. The array has a dimension of 3.7 km. When a muon has an energy of 3 TeV, its trajectory length reaches the size of the array.}
    \label{fig:muon_prop_len}
\end{figure}

As shown in Figure~\ref{fig:muon_prop_len}, the secondary muons with energies exceeding 1 TeV can travel a distance from 1 to 10 kilometers.
Therefore, a detector array with a dimension of kilometers is appropriate for capturing the entire muon track.
\replaced{Above }{Below} this scale, a larger array will intercept a longer muon track, enhancing the precision of directional reconstruction.
However, increasing the size does not appreciably improve directional resolution beyond this dimension.

\begin{figure}[!htb]
    \centering
     \includegraphics[width=0.45\textwidth]{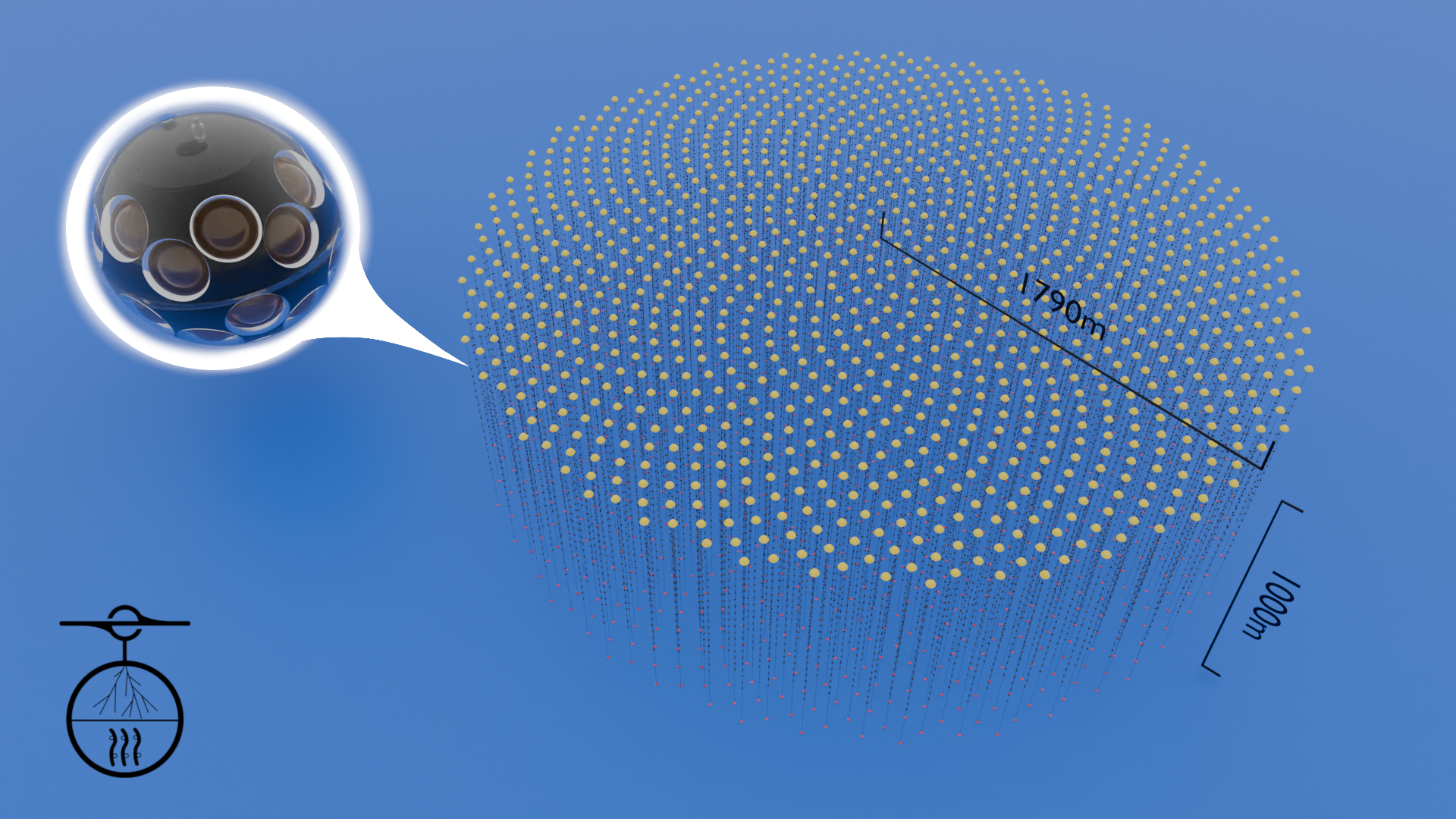}
     \includegraphics[width=0.45\textwidth]{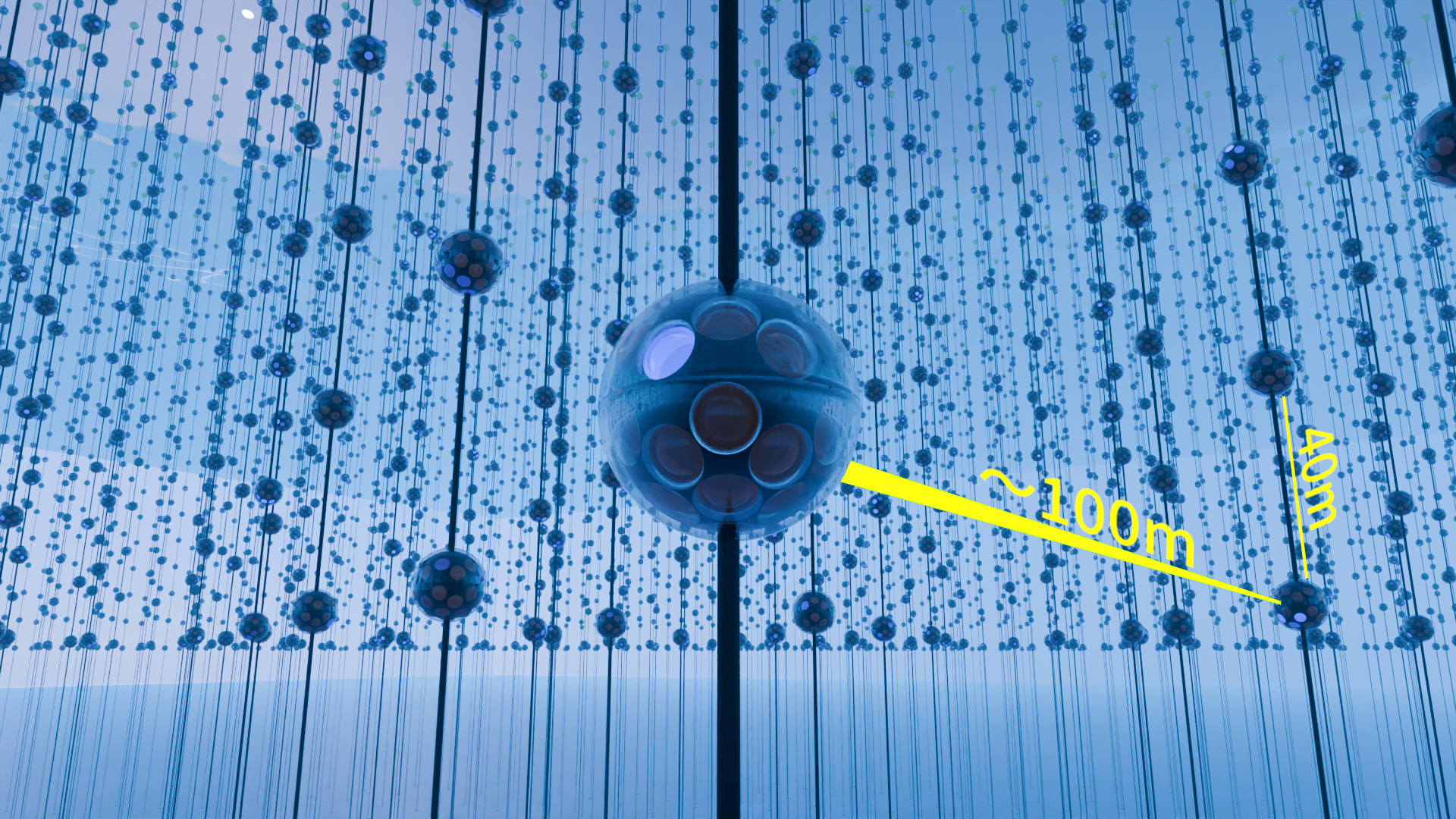}
    \caption{\replaced{The upper panel is a top view of the array. The array is a cylinder with radius of 1.79 km and height of 1 km. The vertical black lines are strings attached with OMs. The red dots at the bottom of the strings represent the anchors, and the yellow dot at the top represent the buoys. The lower panel is a }{A} view of the array from the interior perspective. The OMs are spaced horizontally at approximately 100 meters, with a vertical separation of 40 meters.}
    \label{fig:total_array_perspective}
\end{figure}

It was demonstrated that regular grid layouts, including equilateral triangular and square meshes, may fail to detect muon tracks that coincidentally lie between two columns of strings.
This issue is known as the ``corridor effect''~\cite{aartsen2021icecube}.
To address this issue, we have adopted the Fibonacci spiral~\cite{mungubaComplexBuildAlgorithm2021}, which maintains irregularity while preserving uniformity, to arrange the string of our detector array.

\deleted{The Fibonacci spiral layout in a polar axis can be represented as
\begin{equation*}
    \left\{(\theta_n,r_n):\left(2\pi \left(\frac{n}{\phi} \% 1\right), R\sqrt{\frac{n}{N}}\right) \forall 0 \le n \le N \right\},
\end{equation*}
where the parameter $\phi$ is the Golden Ratio, $R$ is the maximum radius of the array, and $N$ is the total number of points.}

\added{Although the Fibonacci spiral layout is suitable for the detector, it might be not easy in sea operations. We are conducting engineering study to evaluate the feasibility of the realisation of such geometry.} A comprehensive view of the entire array is presented in Figure~\ref{fig:total_array_perspective}.

\subsection{Optical noise filtering} \label{subsec_optical_noise}

In the deep sea environment, there are two
main natural light contributions, bioluminescent and radioactive isotopes. They generate random noise hits that can mislead the event reconstruction.
Bioluminescence originates from a variety of species, spanning from microorganisms to marine life such as fish that inhabit the seafloor.
Although successional  bioluminescent bursts can be eliminated through the implementation of a time cut, random individual bursts constitute the predominant source of bioluminescent noise.
The KM3NeT collaboration has characterized the baseline bioluminescent noise rate with the \added{prototypes of} multi-PMT OM \added{deployed at the site of ANTARES}, observing a rate ranging from approximately 100 kHz to 1 MHz, showing a gradual change over time~\cite{KM3NeT:2014wix}.

The primary source of radioactive emissions in seawater is $^{40}\mathrm{K}$, an nuclide  with a half-life similar to the age of Earth.
This lifetime guarantees that $^{40}\mathrm{K}$ maintains a tolerable level of radioactivity and has not completely decayed throughout the Earth’s history.
$^{40}\mathrm{K}$ decays primarily via beta decay ($^{40}\mathrm{K} \to ^{40}\mathrm{Ca} + e^{-} + \bar{\nu}_e$), or electron capture ($^{40}\mathrm{K} + e^{-} \to ^{40}\mathrm{Ar} + \gamma + \nu_e$), both finally resulting in \deleted{MeV} electrons (positrons) that produce Cherenkov light in water.

The OM is the basic detection unit aligned in strings. 
We adopted the multi-PMT design in order to mitigate the optical noises effectively.
Motivated by the previous work by KM3NeT and IceCube-Gen2\cite{Lohner:2009zz,Kappes:2016kbd}, the designs incorporating 12, 19, 28, and 31 3-inch PMTs are examined in this proposal. 
The multi-PMT design  has a few advantages, which have been thoroughly examined in prior KM3NeT research.
It has a homogeneous photocathode coverage, eliminates the requirement for magnetic shielding, and possesses the ability to differentiate between single-photon and multi-photon hits. All the above features significantly improve the performance of event reconstruction.

By utilizing the multi-PMT configuration, we are able to mitigate optical noise by examining coincident hits within a single OM.
Assuming the optical depth of seawater as 28 meters, our simulation found that $^{40}\mathrm{K}$ yields a noise rate of 110 kHz per OM.
Unlike bioluminescence, $^{40}\mathrm{K}$ noise is consistent and can be measured by analyzing the coincidence rates among multiple PMTs within one OM~\cite{KM3NeT:2014wix}.
We developed an effective filtering strategy to eliminate a substantial portion of these noises \added{with the triggered data}.

\added{In the triggering process, the detector array is divided into multiple overlapping blocks, each with a radius of 100 meters. One block is triggered when exceeding a threshold of 0.6 hits per OM over a 600-ns time window. Only hits recorded within activated blocks are preserved.}
\added{
This triggering algorithm is solely appropriate for offline applications in on-shore infrastructures.
In off-shore facilities, the algorithm must be adapted to accommodate hardware constraints such as bandwidth limitations.
An online counterpart possesses the potential to achieve comparable efficiency.
The online version of this algorithm is currently under development.
}

\replaced{The subsequent filtering process involves the following sequential steps: }{The filtering process comprises the following steps:}
\begin{itemize}
    \item [1.] Exclude hits with a photon electron (pe) number below 0.3.
    \item [2.] Exclude OMs with a low hit count, except when the pe of their hits is large enough.
    \item [3.] Identify peaks \added{($t_p$)} in the hit time distribution for each OM \added{and calculate the standard deviation of hit time relative to its peak ($\sigma_t=\sqrt{\frac{\sum^{i=n}_{i=0} (t_i-t_p)^2}{n}}$).}
    \item [4.] Exclude hits with a deviation time from the peak greater than \replaced{$\max(2\sigma_t, 400\mathrm{ns})$ }{$\max(\sigma_t, 400\mathrm{ns})$}, except when their pe is large enough.
    \item [5.] Repeat steps 2 to 4 until \replaced{the $\sigma_t$ is acceptably (<200ns) small or remains unchanged. }{the standard error of the residual hit times is acceptably small.}
    \item [6.] Conduct a linear fit in space-time for the remaining hits and exclude those that deviate significantly from the fitted line.
\end{itemize}

In Step 2, unrelated hits are effectively filtered out, which enhances the efficiency of the subsequent stages.
Steps 3 and 4 involve identifying time-correlated hits originating from the Cherenkov radiation cone through a time distribution analysis, thereby excluding noise hits that either arrive significantly before or after these hits.
After the iterative process in Step 5, Step 6 utilizes linear regression for a rough reconstruction of the muon track and excludes hits that deviate substantially from the track.

\begin{figure}[!htb]
    \includegraphics[width=0.45\textwidth]{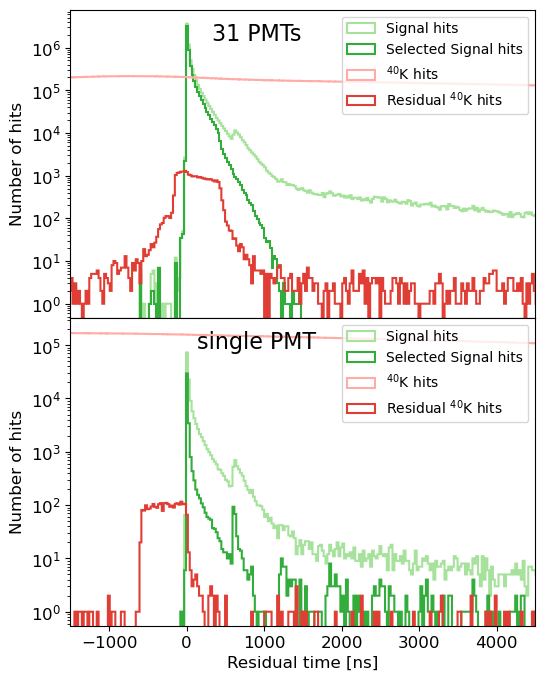}
    \caption{Signal and noise hit distributions as a function of residual time, illustrating the effects of filtering. Green and red lines denote signal hits and homogeneous $^{40}$K hits, respectively. Lighter and darker colors distinguish between distributions before and after filtering. The upper panel presents data from the 31-PMT configuration, and the lower panel presents data from the single-PMT configuration.}
    \label{fig:allvssel}
\end{figure}

We demonstrate the efficiency of this process by examining the residual time distribution of hits. As depicted in Figure \ref{fig:allvssel}, the 31-PMT configuration is chosen as a representative case for comparison with the single-PMT configuration.The signal hits and homogeneous $^{40}$K hits are shown with light green and light red lines, respectively.
Following the filtering process mentioned above, the processed hits are presented with corresponding dark green and dark red lines, respectively.
In the case of multiple-PMT configurations, the algorithm significantly reduces noise hits by $\sim$4 orders of magnitude with a signal preservation rate of approximately $78\%$.
Around the Cherenkov radiation cone, \replaced{the algorithm reduces noise to about $10^{-3.5}$ of the signal. }{the algorithm achieves a noise reduction of 3.5 orders of magnitude relative to the signal.}
In contrast, the filtering performance for a single-PMT configuration is also presented in the lower panel.
In this case, the aforementioned algorithm is no longer available as there are no coincident hits within a single OM.
Instead, a PE threshold of 5 is used to replace the algorithm.
The absence of coincident hit data weakens the noise-signal discrimination, resulting in the preservation of roughly $26\%$ of the signal and \replaced{noise is reduced to only $10^{-2.5}$ of the signal. }{a noise decrease of only 2.5 orders of magnitude in the vicinity of the cone.}

\deleted{The filtering algorithm discussed is solely appropriate for offline applications. Nevertheless, an online counterpart possesses the potential to achieve comparable efficiency. The online version of this algorithm is currently under development.}

\subsection{Detector development} \label{subsec:detector_development}

In order to finally implement an appropriate multi-PMT design, we are currently developing a preliminary version of OM and performing hardware tests on it.

The onshore tests are conducted  on two brands of 3-inch PMTs, the NVN N2031 and the Hamamatsu R14374, to evaluate their quantum efficiency, gain, and dark count rate.
Both PMTs demonstrate similar performance, although the Hamamatsu R14374 was found to have superior quality control.
The quantum efficiency for both PMT models spans from 25\% to 32\%.
\remove{When operated at voltages between 950V and 1050V,}
The NVN N2031 and the Hamamatsu R14374 PMTs operate within voltage ranges of 950V to 1200V and 950V to 1050V, respectively.
Both PMTs exhibit gains of approximately $5 \times 10^6$ and dark count rates ranging from 200Hz to 3kHz.
Additional onshore tests are currently being conducted on the assembled OM using the two types of PMTs. The detailed results will be presented in our separate, upcoming works. 

In the spring of 2024, we conducted an offshore test of a preliminary version of OM in the Qiandao Lake, which validated the functionality of our electronic systems.
Currently, the OM is still under further development.

We plan to integrate a hierarchical data collection system into our array.
The array is divided into multiple clusters, each extending horizontally approximately 100 meters.
A junction box situated within each cluster is responsible for processing all detected hits.
Subsequently, only hits triggered at the cluster level are relayed to the central junction box during subsequent passes, where they are merged into events.

The hierarchical data collection system enables us to initiate data acquisition in a hierarchy manner.
Considering that a high-energy secondary particle is capable of triggering the OM only within a radius of $\sim$100 meters, we have divided the array into subblocks, each with a size of 100 meters.
These subblocks are aligned with the previously mentioned string clusters to simplify their management.
We trigger subblocks that meet or exceed a determined total hit count threshold, thereby excluding those subblocks that do not register any signals.
Following this, the triggered subblocks transmit their hit data to a central junction box.
The central junction box then merges the data from the subblocks that have been triggered close in time.

In this proposal, this trigger algorithm is subsequently applied to simulation data to facilitate typical estimations.

\section{NEON simulation framework}\label{sec:simulation}
\begin{figure*}[!htb]
    \centering \includegraphics[width=0.8\textwidth]{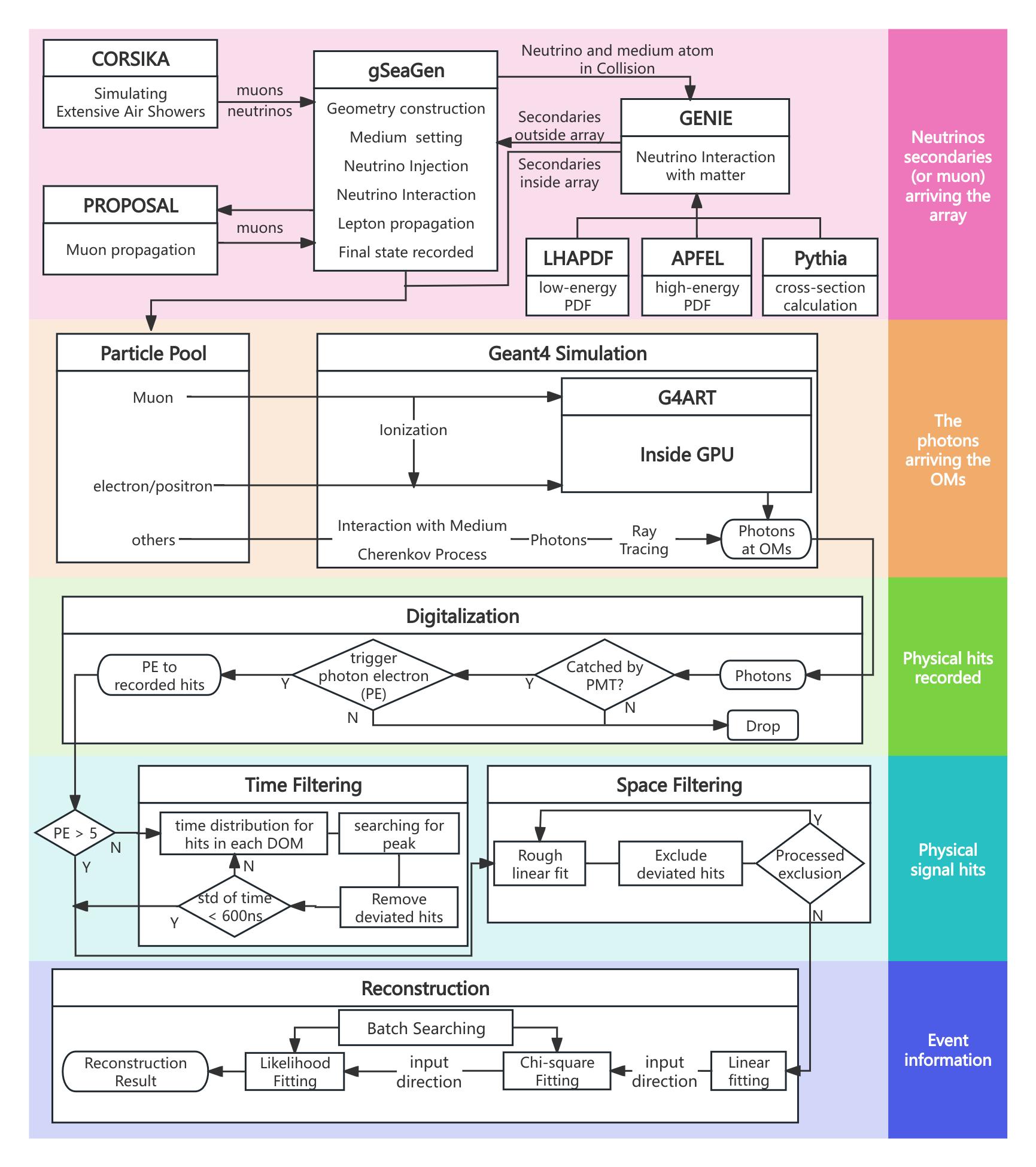}
    \caption{Flowchart of the NEONSim and the reconstruction.}
    \label{fig:Simulated_flowchart}
\end{figure*}
Monte Carlo simulations are employed to study the detector's response to both neutrino signals and background events. The simulation framework for NEON, so-called NEONSim, consists of four parts. 

The first part simulates the flux of atmospheric muons and neutrinos by modeling the propagation of high-energy air showers, which are triggered by cosmic rays.
The second part calculates the interactions of neutrinos with nucleons within seawater, resulting in the generation of detectable secondary particles.
The third part evaluates these secondaries and the Cherenkov photons emitted by them within the detector array.
The fourth part converts the recorded Cherenkov photons into physical hit signals.

\subsection{Atmospheric muon and neutrino background}
\label{subsec:atmospheric_muon_and_neutrino_background}

\begin{figure*}[!htb]
    \centering
    \includegraphics[width=0.9\textwidth]{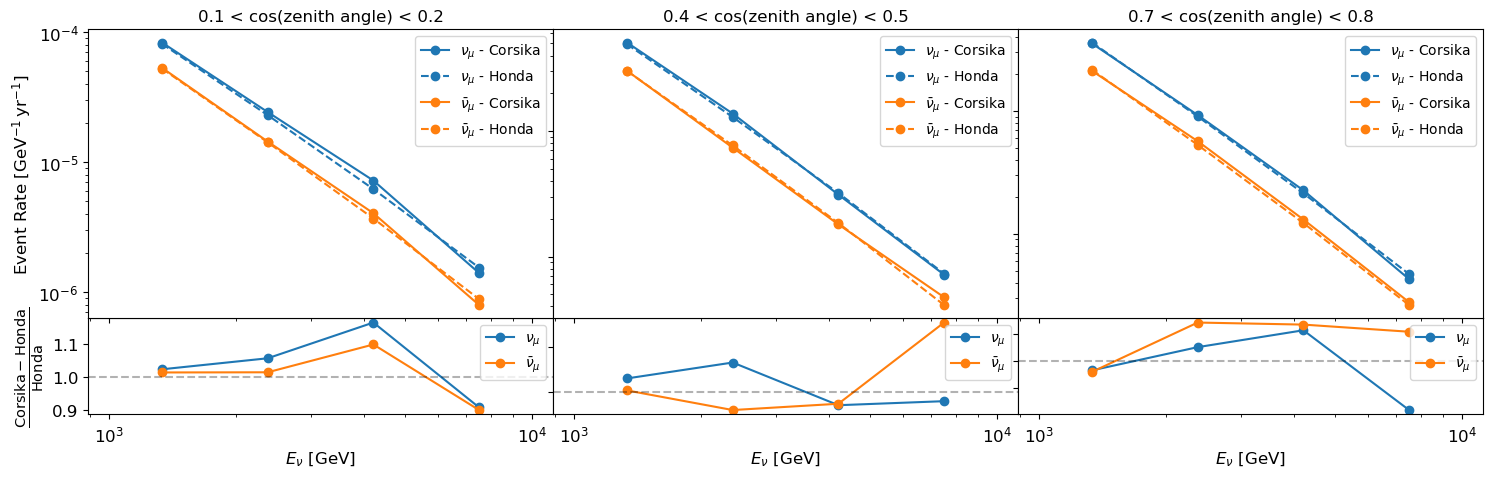}
    \caption{Comparison between simulated atmospheric neutrino (antineutrino) flux and Honda model, with various $\cos(\theta_\mathrm{zen})$ ranges.}
    \label{fig:atmonu}
\end{figure*}

The main backgrounds for astrophysical neutrino telescopes are atmospheric muons and neutrinos, both originating from extensive air showers (EAS) triggered by high-energy cosmic rays.
These particles result from the decay of mesons, such as $\pi$ and $K$.
The simulation of the EAS process is conducted with the CORSIKA framework~\cite{Reininghaus:2019jxg}.
In the simulation, the high-energy cosmic ray flux model proposed by Gaisser et al.~\cite{Gaisser:2013bla} serves as the input.
We compare our simulated atmospheric neutrinos to the widely used results from the model presented by Honda et al.~\cite{Honda:2015fha}, as illustrated in Figure~\ref{fig:atmonu}.
Our predictions for atmospheric neutrinos and antineutrinos in the energy range from $1\TeV$ to $10\TeV$ are in good agreement with Honda's results and are naturally extended to the high-energy region up to $100\PeV$.

In addition to the decays of light mesons mentioned above, prompt muons and neutrinos can also arise from the decays of heavy mesons containing heavy quarks as valence quarks, including $D$ and $\Lambda_C$.
These heavy mesons decay promptly, as their decay lengths are shorter than their interaction lengths.
At high energies, the prompt atmospheric neutrino flux becomes dominant, surpassing the conventional one across energy from tens of $\TeV$ to $\PeV$, with the variations according to the models~\cite{Jeong:2023gla, Fedynitch:2015zma}.
However, this effect is currently beyond the sensitivity of neutrino telescopes such as IceCube. A recent analysis using IceCube's starting track events has only established an upper limit on its presence~\cite{IceCube:2024fxo}. 

We plan to integrate this prompt contribution into our simulation processes in the future.
Currently, we estimate this contribution through a scaling operation, with the scaling factor derived from the results of Ref.~\cite{Jeong:2023gla}.

Atmospheric neutrinos can be directly injected into the array.
However, atmospheric muons experience energy loss as they traverse through seawater.
The PROPOSAL~\cite{Koehne:2013gpa} package is employed to simulate the energy loss of these muons.
The energy loss of atmospheric muons is affected by the distance they travel through seawater, which is determined by the zenith angle $\theta_\mathrm{zen}$ and the depth of the observation point.
Based on the PROPOSAL, we developed a program that propagates muons in a stepwise manner, allowing us to acquire the distribution at multiple depths from a single simulation.

The vertical muon flux has been measured at various water depths by several neutrino telescope experiments.
We have simulated the muon flux at different depths using our program based on the PROPOSAL code, and have compared these simulations with the experimental data.

\begin{figure}[!htb]
    \includegraphics[width=0.45\textwidth]{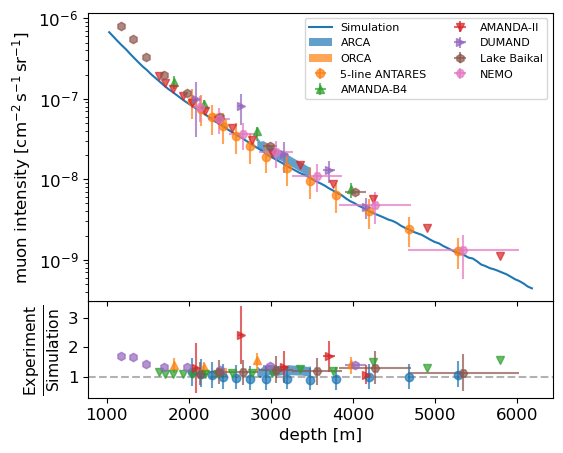}
    \caption{Comparison between simulation results of atmospheric muon vertical flux \added{(>1GeV)} and measurements from previous experiments, DUMAND\cite{Clem:1990iz}, Lake Baikal\cite{BAIKAL:1997iok}, AMANDA\cite{Bai:2003mn}\cite{Andres:1999hm}, NEMO\cite{NEMO:2009vtd}, ANTARES\cite{ANTARES:2010izk}, ORCA and ARCA\cite{KM3NeT:2019jfa}. The results of ORCA and ARCA utilized the parameterized equations provided in the article to obtain the vertical atmospheric muon flux\cite{KM3NeT:2019jfa}.}
    \label{fig:atmomuon}
\end{figure}

As illustrated in Figure~\ref{fig:atmomuon}, our simulation  is compatible with the experimental measurements.

\subsection{Neutrino Interaction}

The weak interaction is a well-established theory. However, computing neutrino collisions with nuclei encounters significant challenges. 
In these interactions, neutrinos directly interact with the quarks within the nucleons. A comprehensive simulation of such collisions necessitates the integration of the target nucleon's Parton Distribution Functions (PDFs) and the interactions of the resulting secondary particles with the other nucleons within the nucleus.

The Generates Events for Neutrino Interaction Experiments (GENIE) project~\cite{Andreopoulos:2009rq} was established to address the complexities mentioned above, providing comprehensive physics models for simulating neutrino interactions.
It currently supports the generating of events with energies up to multi-TeV and provides a selection of comprehensive model configurations, each with distinct tuning calibrated against various experimental data.
We utilize the GHE19\_\-00b\_\-00\_\-000 tuning for our simulations.
To explore higher energies, we have extended the cross-section up to 100 PeV, utilizing the tools provided by the GENIE/Generator package alongside PDFs calculated by APFEL~\cite{Bertone:2017gds}.

Based on the GENIE project, the code gSeaGen~\cite{KM3NeT:2020tvi} was developed to simulate detectable events surrounding a neutrino telescope.
For events occurring outside the array, the PROPOSAL package~\cite{Koehne:2013gpa} was adopted to simulate muon propagation through the material.

\subsection{Cherenkov Photon Signal}

With the neutrino interaction products generated by gSeaGen, we performed detailed simulations using Geant4 \cite{agostinelli2003geant4}.
High-energy muons lose energy through ionization, Compton scattering, and pair production, producing high-energy electrons and positrons as they pass through water.
Subsequently, Cherenkov photons are predominantly emitted by these produced electrons and positrons with speed exceeding the light speed in seawater, corresponding to a kinetic energy threshold of 236 \replaced{keV }{eV}.
These photons propagate through water and undergo scattering before being detected by PMTs or absorbed.

In Geant4 simulations, the vast number of photons consumes a majority of the computing resources.
To overcome this difficulty, KM3NeT employs direct sampling of hits to economize on this consumption, while IceCube utilizes the GPU for their photon propagation. The HUNT project software team led by Shun Xu has developed the G4ART\cite{G4ART} software package. We validated the G4ART in the NEONSim framework.
G4ART enables parallel processing of photons on the GPU, simulating their generation, translation, absorption, and scattering.
\remove{To prevent the time-consuming process of copying photons from Geant4's memory to the GPU, G4ART generates photons directly on the GPU.
This approach has been shown to be 10 times more efficient than directly copying the photon during the initial development stages of G4ART.}
The latest version of G4ART has been demonstrated to exhibit an efficiency approximately 200 times greater than the original Geant4.

\remove{
High-energy charged particles passing through seawater produce numerous high-energy electrons and positrons, which then initiate electromagnetic cascades.
The Cherenkov photons emitted from all these particles form a photon shower.
Re-implementing these complex cascade and Cherenkov processes entirely on a GPU would necessitate considerable effort.
Therefore, G4ART employs a compromise approach.
For particles with energies over 100 GeV, G4ART allows simulations to continue within Geant4 and only captures essential step information to generate their Cherenkov radiation on the GPU.
In contrast, for electrons and positrons with energies below 100 GeV, G4ART directly captures them and calculates the resulting photon showers on the GPU through parameterization.
}

\remove{
It is estimated that high-energy particles produce approximately 1000 Cherenkov photons per step in Geant4, while an electron or positron with an energy of around 100 GeV results in a shower containing $\sim 10^7$ Cherenkov photons.
Transferring step and particle information directly from memory to the GPU can avoid the necessity of copying these photons.
}

\replaced{The information pertaining to high-energy charged particles within Geant4 is transferred to G4ART. G4ART then utilizes this information to generate Cherenkov photons directly within the GPU's memory. }{G4ART selectively captures the trajectory of high-energy charged particles or the particles themselves, rather than the Cherenkov photons.}
This approach avoids the time-consuming task of transferring large quantities of photons from memory to the GPU.

\remove{Electromagnetic cascades with energy scales around 100 GeV produce a sizeable number of secondary particles.
Consequently, these cascades exhibit symmetry and stability, which allows for their parameterization.}
In order to simulate the high-energy particles inside GPU, we have developed a parameterization for the distribution of shower photons emitted by the induced electromagnetic cascades, accounting for their initial position, starting time, direction and energy.
After fitting the parameterization model \cite{radel2013calculation} in our simulation with original Geant4, we implement the parameterization through the methods provided by G4ART. In the parameterization model,
the initial positions of these photons concentrate around a path originating from the initial particle and extending along the direction of momentum, over a distance of several meters.
This distribution would vary with the energy of the initial particle.
The photon starting time is approximately the distance from the initial particle's position divided by the speed of light. 
The angular distribution of photons to the direction of the initial particle's momentum is parameterized using the function
\begin{equation}
\dfrac{\mathrm{d}n}{\mathrm{d}\Omega}(\theta) = a \exp\left({b\left|\cos\theta - \cos \theta_\mathrm{Che}\right|^c}\right) + d,
\end{equation}
where a, b, c, and d are free parameters, $\theta_\mathrm{Che}\approx 41^\circ$ is the Cherenkov angle in water~\cite{radel2013calculation} and $\theta$ is the angle between initial particle and photon.
The energy spectrum of these Cherenkov photons is described by the Frank–Tamm formula~\cite{vcerenkov1937visible}, incorporating the energy-dependent refractive index of water.

The primary distinction between complete simulation and parameterization method is the simplification of the initial spatial-momentum distribution of photons. This simplification may result in slight deviations in the number of photons detected by the OMs. To validate the parameterization method, we compare the number of photon hits to each OM with and without G4ART for neutrino events.

\begin{figure}
    \centering
    \includegraphics[width=1\linewidth]{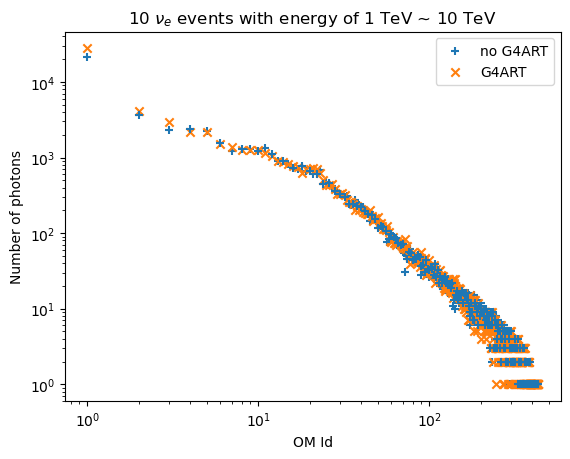}
    \caption{The figure shows the number of photon hits to each OM with (blue plus sign) and without G4ART (orange cross sign) in 10 events. The OM Ids are sorted by the number of hits.}
    \label{fig:G4ART_validation}
\end{figure}

As shown in Figure \ref{fig:G4ART_validation}, we concatenate the results from 10 cascade events, sorting $N_\mathrm{photon}$ in descending order.
The results obtained with and without G4ART show a good agreement.

By utilizing the G4ART framework, we can efficiently accumulate high-energy events for both muon and neutrino injections.
Simulating the Cherenkov photons from these events and analyzing the PMT signals produced by them allows us to assess the performance of the detector array.

We employ Geant4 to track all photons that strike the surface of the OMs.
However, these photons do not represent physical hits.
A hit is recorded only when a photon enters a PMT, excites a photoelectron, and ultimately triggers a multiplication event in the PMT's multiplication region.
Subsequent photons that arrive with a time separation of less than the dead time will be merged into one single hit.
Taking these processes into account, we generate hit records for the 12-PMT, 19-PMT, 28-PMT, and 31-PMT configurations of the OMs.

\section{Detector Performance}\label{performance}

In this section, we present the performance studies of NEON, including the angular resolution, effective areas, and sensitivities to the point-like sources.
At the current stage of analysis, only muon (anti-)neutrinos ($\nu_\mu$/$\bar{\nu}_\mu$) are considered.

\subsection{Event reconstruction}

Event reconstruction aims to determine the trajectory direction.
A muon's trajectory involves five independent parameters, including direction and vertex coordinates.

Cherenkov photons are emitted either by the muon itself or by the electrons and positrons induced.
As the induced electrons and positrons retain a directional alignment with the original muon, all these Cherenkov photons will diverge from the trajectory at an angle approximately equal to the Cherenkov angle in seawater.
By determining the precise relationship between the time and position of these photons in such a scenario, a chi-square test can be constructed as $\chi^2\equiv\sum_i (t_\mathrm{det}^i - t_\mathrm{the}^i)^2$ to compare the theoretical predicted time $t_\mathrm{the}^i$ with the actual time recorded by OMs $t_\mathrm{det}^i$.

Due to the photon scattering in seawater, the recorded time $t_\mathrm{det}^i$ are systematically delayed relative to the predictions $t_\mathrm{the}^i$.
This results in a long tail in the distribution of the residual time $t_\mathrm{res}^i \equiv t_\mathrm{det}^i - t_\mathrm{the}^i$. 
To account for this asymmetric distribution of residual times, we have also implemented a likelihood fitting with $\mathcal{L} \equiv \Pi_i f(t_\mathrm{res}^i)$, where the distribution $f(t)$ is adopted from the simulation results. 

Directly scanning the entire parameter space to identify the most likely parameter is both time-consuming and unstable due to the presence of numerous local minima, particularly when the number of hits is insufficient.
Consequently, an iterative batched local search algorithm is implemented. 
In the search, the entire sky is divided into equal-area pixels according to the HEALPix partitioning method \cite{gorski2005healpix} .
At first, a reference direction, anticipated to be roughly aligned with the true direction, is determined.
Subsequently, a circular searching window of radius $R$ centered on the reference direction is established.
A batch of local optimizations is performed, starting from the center of each pixel within the window. 
The optimization result with the maximum likelihood is selected as the new reference direction, upon which a new batch of searches is initiated.
The algorithm stops when the fitting result no longer improves or the angle between the reference directions from two consecutive iterations is less than $0.1^\circ$.

This batch searching ensures that the optimization result is the true best fit within the circular window, even when there are numerous local minima.
The iterative process allows the reference direction to shift towards the true global direction, which may lie outside the circular window.

In the practical implementation, a linear fit of the recorded hits is initially performed to roughly determine the muon direction.
This direction serves as the input reference direction for an iterative batched search to find the minimum $\chi^2$ value. 
The search algorithm utilizes a search window with a radius of $R = 10^\circ$ and a HEALPix dividing order of $N_\mathrm{order} = 5$, which corresponds to pixel sizes of approximately $2^\circ$.
Subsequently, an additional iterative batched search for the maximum $\mathcal{L}$ is conducted using a window radius of $R = 2^\circ$ and a HEALPix order of $N_\mathrm{order} = 7$, corresponding to pixel sizes of about $0.5^\circ$, to further refine the reconstruction accuracy.

To assess the quality of event reconstruction, a quality parameter $\Lambda$, defined as $- \log(\mathcal{L})/(N_\mathrm{hit}-5)$, is introduced.
Events with lower $\Lambda$ values typically exhibit smaller reconstruction errors, $\Delta\Omega$, which represent the angular discrepancy between the true and reconstructed muon trajectory directions.
A $\Lambda$ threshold is employed to reject poorly reconstructed events.

We establish a $\Delta\Omega$ threshold of $1^\circ$ to differentiate between `true' and `false' events. 
The True Positive Rate (TPR) and False Positive Rate (FPR) are calculated using the following equations: \begin{equation}
\mathrm{TPR}\equiv\frac{\mathrm{N}^{\mathrm{cut}}_{\Delta\Omega < 1^\circ}}{\mathrm{N}_{\Delta \Omega < 1^\circ}},\\
\mathrm{FPR}\equiv\frac{\mathrm{N}^{\mathrm{cut}}_{\Delta \Omega > 1^\circ}}{\mathrm{N}_{\Delta \Omega > 1^\circ}}
\end{equation}
Where N is the number of neutrino events reconstructed successfully\added{, and $\mathrm{N^{cut}}$ represents the subset of these events with $\Lambda$ below a cut of 3.9}. These rates serve to evaluate the efficiency of event rejection.
By setting a $\Lambda$ cut at $3.9$, we maximize the value of $\rm TPR-FPR$, which leads to the retention of approximately $76\%$ of good events and the rejection of about $64\%$ of poor events.

\begin{figure}[!htb]
    \includegraphics[width=0.45\textwidth]{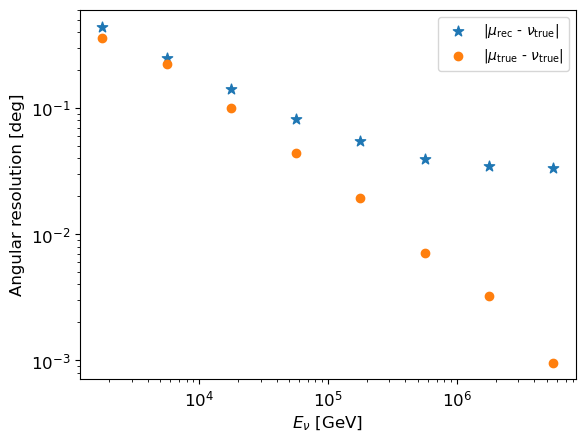}
    \caption{The median angular error is plotted as a function of the neutrino's true energy. Orange circles represent the angle between the muon track and the neutrino's true direction, while blue stars denote the angle between the reconstructed track and the neutrino's true direction. }
    \label{fig:angularres}
\end{figure}

Following the application of a cut to retain good events, we compute the angular resolution across various energy bins, as depicted in Figure \ref{fig:angularres}.
The angle between the muon trajectory and the original muon neutrino decreases with energy, following a power-law.
Below $\rm 100\ TeV$, this systematic error predominates in the overall directional resolution.
However, at energies exceeding $\rm 100\ TeV$, the muon track reconstruction error, which is roughly $\rm 0.05^\circ$, becomes the dominant factor.

\subsection{Sensitivity Analysis}

With the simulation and reconstruction of background and signal events, the effective area and sensitivity of the NEON experiment can be obtained.
In the signal simulations, we inject 452732 weighted $\nu_\mu$ (${\bar{\nu}}_\mu$) events, spanning from 1 TeV to 100 PeV and uniformly distributed over solid angle into the detector array.
Of these, 370000 events are triggered, about 48000 events has signal hits, and about 10000  are successfully reconstructed.

The current version of the trigger algorithm is somewhat loose, resulting in the inclusion of excessive noise-induced events in the analysis.
The trigger threshold will be adjusted again to address this issue once the noise hit rate is measured.

Utilizing this dataset, we calculated the effective area for the 12-PMT, 19-PMT, 28-PMT, and 31-PMT configuration separately.
Notably, the effective area decreases with the decrease in the number of PMTs at lower energy ranges.
The 19-PMT configuration maintains an effective area that is comparable with the 31-PMT configuration across the energy range of 1 TeV to 10 TeV.
The 19-PMT configuration appears to strike a balance between performance and power consumption.
Nonetheless, when the power supply is not limited, the 31-PMT configuration is preferable.
In the following analysis, we choose 31-PMT configuration as a typical case.

\begin{figure}[!htb]
    \includegraphics[width=0.45\textwidth]{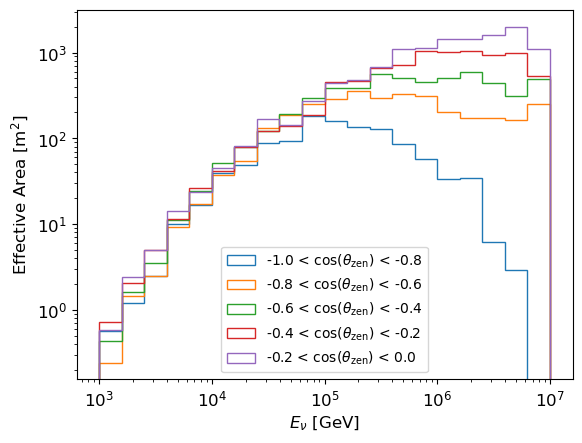}
    \includegraphics[width=0.45\textwidth]{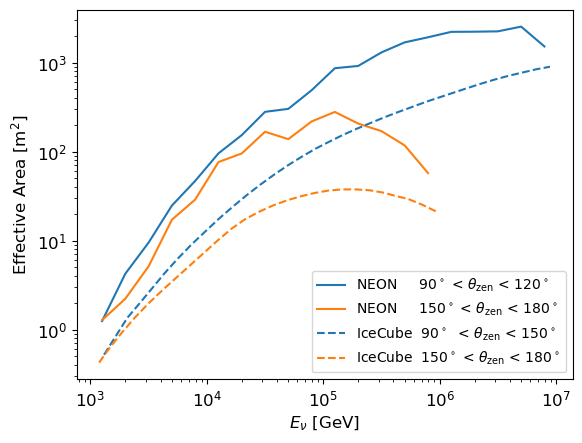}
    \caption{The upper panel shows the average effective areas for various zenith angles and energies.
    The lower panel compares the effective areas of NEON and IceCube for two zenith bin of 90$^\circ$ -- 120$^\circ$ and 120$^\circ$ -- 150$^\circ$ \cite{navas2024review}.
    }
    \label{fig:effareas}
\end{figure}

Figure~\ref{fig:effareas} illustrates the effective area as a function of varying zenith angles and energies for 31-PMT configuration.
In the high-energy region, the effective area reaches its maximum in horizontal directions as neutrinos from these paths can evade both Earth's absorption and the muon background interference.
Typically, the horizontal effective area reaches an area of $\added{\sim}10^3$ m$^2$ at the energy of PeV.

As discussed in Section \ref{subsec:atmospheric_muon_and_neutrino_background}, the primary background events in this analysis are atmospheric muons and neutrinos.
As the atmospheric \added{muon} neutrino behaves the same as astrophysical \added{muon} neutrinos, the simulation of muon events is conducted as part of the analysis process.
Unlike neutrinos, almost all muons with sufficient energy  passing through the array would trigger the OMs. 
Consequently, the effective area for muons is roughly 4 orders of magnitude greater than that of neutrinos at TeV energies.

The muon neutrinos that collide with atoms within the array always produce tracks with origins located inside the array, known as starting tracks.
Conversely, high-energy atmospheric muons or muon neutrinos that collide with atoms outside the array generate tracks that traverse the array, referred to as through-going tracks.
Consequently, distinguishing between through-going tracks and starting tracks can filter out the background muons.
Following the filtering of hit noises, the identification is made through the following steps:

\begin{itemize}
    \item [1.] \textbf{Time and Weight Assignment:} The average time of all hits in each OM is utilized as the trigger time for OM. Each OM is then assigned a weight number $w$ in following the rules
\begin{equation*}
    w = \left\{\begin{array}{ll}
        1  &  N_{hit}<15 \\
        2  &  15\le N_{hit} < 30\\
        3  &  30 \le N_{hit}
    \end{array}\right.;
\end{equation*}
    \item [2.] \textbf{Calculation of the ``Invisible Trace" Length:} The length $L$ of the ``invisible trace" (in meters) is determined by the length of the muon track from the earliest triggered OM to the subsurface layer; 
    \item [3.] \textbf{Calculation of the Starting Counts:} For the events with a triggered time window exceeding 900 ns, the sum of the weighted number of OMs triggered within the initial 600 ns is calculated and recorded as the starting counts $N_{start}$;
    \item [4.] \textbf{Event Differentiation:} Events are classified within the plane defined by $ L $ versus $ N_{start} $. Only events satisfying the condition $ L > 100 $ and  $ N_{start} > -\frac{L}{600} + 5 $ are retained.
\end{itemize}

\added{As the hits have been filtered, the timing of hits on OM is focused within a 1000 ns time window. }In Step 1, we use the average timing of all signal hits to determine the trigger time for the OM, and assign a weight $w$ to the OM based on \replaced{the total number of hits on OM $N_{hit}$. }{the number of hits $N_{hit}$.}
We prefer $N_{hit}$ over the total number of photon electrons, $N_{pe}$, as the latter is significantly influenced by the proximity between the track and the OM.
The weight factor $w$ is constructed to maintain stability: Cherenkov light from muons always yields $w=1$, hadronic cascade light can activate more than half of the PMTs to achieve $w=2$, and only OMs in close proximity to the core of a cascade are almost entirely triggered, reaching $w=3$.
In Step 2, we account for the likelihood that through-going particles will trigger the array upon entry.
To this end, we introduce the parameter ``invisible trace length'' $L$ in our discrimination process.
We expect $L$ to be minimal for through-going events, particularly in high-energy regions.
Consequently, events with a significantly long $L$ are likely the starting tracks.
In defining $L$, we select the subsurface layer rather than the outermost layer in case through-going events are undetected by the surface layer. For our array layout, OMs with a vertical distance of no more than 40m or a horizontal distance of no more than 100m from the surface layer are considered as the "subsurface layer".
In Step 3, we utilize the counts of OMs within the initial 600 nanoseconds, $ N_{start} $, to represent the energy deposition around the starting point.
Starting tracks typically dissipate their cascade energy near the origin, resulting in a higher $N_{start}$ value compared to through-going tracks.
By investigating the $ L $ versus $ N_{start} $ plane, we finally achieve the discrimination.

Following this discrimination, only $10^{-4}$ of the muon remains, while $55\%$ of the initial neutrino from the starting track event is preserved.
By applying the flux of atmospheric muons and neutrinos simulated in Section \ref{subsec:atmospheric_muon_and_neutrino_background} the background event rate is obtained.
With the assumption of a spectral index of -2 and a minimum energy of $10\TeV$, we calculate the required flux for a $5\sigma$ discovery of a point-like source.
The counting window for the background and the source is set according to the Point Spread Function (PSF) to maintain a signal percentage that maximizes statistical significance.
The accurate angular resolution significantly reduces the background count.

\begin{figure}[!htb]
    \includegraphics[width=0.45\textwidth]{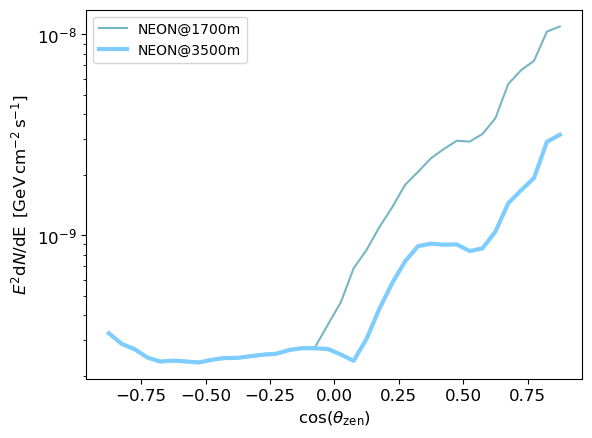}
    \caption{$5\sigma$ discovery potential for NEON-1700m and NEON-3500m of different zenith.}
    \label{fig:5sigma_z}
\end{figure}

Figure~\ref{fig:5sigma_z} illustrates the $5\sigma$ discovery potential as a function of the zenith angle, based on a 10-year statistical time frame.
Both the array depth of 1700 m and 3500 m are analyzed.
The figure illustrates that in the upper hemisphere, the sensitivity of the 3500 m depth is superior to that of the 1700 m depth, primarily due to the decreased atmospheric muon noise at the greater depth.
However, this figure is only meaningful for transient sources.
Steady sources would move with the rotation of the Earth, thus mixing observations from different zenith angles.

\begin{figure}[!htb]
    \includegraphics[width=0.45\textwidth]{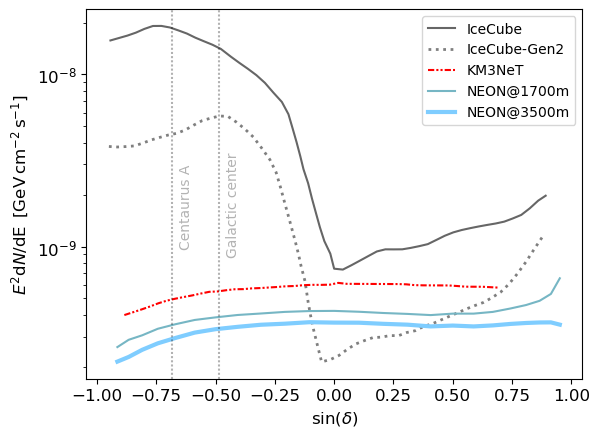}
    \caption{
    The $5\sigma$ discovery potential for NEON-1700m and NEON-3500m, varying by declination based on a 10-year statistical analysis. For comparison, the discovery potentials of IceCube~\cite{IceCube:2019cia,IceCube-Gen2:2021tmd} and KM3NeT~\cite{KM3NeT:2018wnd} are also included, with the results scaled to a 10-year time frame. The declination of Galactic center and Centaurus A  is shown as grey dotted lines.
    }
    \label{fig:5sigma}
\end{figure}

Figure~\ref{fig:5sigma} presents the $5\sigma$ discovery potential for various declinations $\delta$, derived from a 10-year statistical analysis, in comparison with the IceCube~\cite{IceCube:2019cia,IceCube-Gen2:2021tmd} and KM3NeT~\cite{KM3NeT:2018wnd} experiments.
This analysis excludes events with low zenith angles where the count of atmospheric muons exceeds that of atmospheric neutrinos.
The transition from a depth of 3500 meters to 1700 meters does not lead to a significant enhancement in sensitivity.
This is because the discrimination process between through-going track events and starting tracks for the down-going events filters out a significant portion of the signals.
As a result, the total count of the signals is dominated by up-going events, which are atmospheric muon free.
It is necessary to note that this discovery potential analysis is dominated by the events around the energy threshold $10\TeV$.
The up-going events with energy above $\PeV$ would be mostly absorbed by the Earth and the depth advantage of 3500 meters would become more pronounced.

The exclusion of through-going tracks not only removes atmospheric muons but also eliminates the associated atmospheric neutrino background noise.
This process is known as the self-veto of atmospheric neutrinos.
When the algorithm for distinguishing through-going from starting tracks is sufficiently efficient, atmospheric muons can be utilized as a handle to exclude neutrinos.
In the future, we plan to develop these discrimination algorithms further, exploring both more detailed physical methods and machine learning techniques, such as Graph Neural Networks (GNNs). 
If a highly proficient algorithm is available by then, the array at 1700 meters depth could potentially outperform the one at 3500 meters, owing to its greater self-vetoing capability.

Because the location of NEON is in the Northern Hemisphere and close to the equator, NEON is more sensitive to sources in the Southern sky and has a wider field of view. This will be complementary to IceCube. In particular, as seen in Figure \ref{fig:5sigma} the Galactic plane, where Galactic PeVatrons accelerate cosmic rays, is one of the primary targets for NEON. The synergy with gamma-ray observatories, such as CTA and LHAASO, will unveil the mystery of cosmic-ray origin. Moreover, Centaurus A as the closest active galaxy is believed to be a plausible source of ultra-high energy cosmic rays registered by the Pierre Auger Observatory \cite{PierreAuger:2022axr}, so that neutrinos will be produced during the cosmic ray interaction. As shown in Figure \ref{fig:5sigma}, NEON will be able to observe this extragalactic source with the best sensitivity. 

\section{Summary and discussion}\label{sec:summary}

Neutrinos as the unique messenger can bring us information from regions of space where no other photons can reach us. In the last decade, the IceCube collaboration has made several achievements as the first discovery of high-energy astrophysical neutrinos, and identifying neutrinos from single cosmic sources. KM3NeT collaboration found the first neutrino \added{event} with an \replaced{estimated energy of 220 PeV\cite{km3net2025observation} }{energy of tens of PeV\cite{paschal_anthony_coyle_status_nodate}}. All these discoveries have enhanced our knowledge of both neutrino physics and multi-messenger astronomy.  

To date, the two identified neutrino potential sources, TXS 0506+056 and NGC 1068 as two types of AGN, differ in both density and luminosity by orders of magnitude, and each contributes only no more than 1\% to the overall diffuse flux of astrophysical neutrinos in their respective energy ranges. The Galactic diffuse neutrino flux may contribute about 10\%. There might be a large population of unidentified faint sources contributing to the measurements, which demands detectors with better sensitivity. 
Therefore, to fully understand the universe and answer the open questions of physics and astronomy, we propose the next-generation neutrino telescope NEON in this work.

We employ multi-PMT OMs arranged in a Fibonacci spiral within the telescope. 
The test, design, and development of these OMs are in progress.
A simulation framework, NEONSim, has been developed and optimized to manage the simulation process.

With the adoption of GPU, the simulation process has been accelerated by about 200 times.
This speeds up the simulation of events with energies around 10 PeV from approximately 10 days per event to roughly 2 hours per event, making detailed performance analysis of high-energy events feasible.
With the simulated sample, we evaluate the performance of the array with two possible sites, with depths of 1700 and 3500 meters, respectively.
With our background analysis algorithms, the atmospheric muon has been reduced by four orders of magnitude.
\added{We plan to add OMs with different designs to the top of the array, which would  further enhance the veto effect on atmospheric muons. }
A stepwise approach was developed to simulate the propagation of atmospheric muons, enabling the calculation of noise characteristics at various depths simultaneously.
We find for the detector deployed in the deep sea, there is not much matter for depth between 1700 and 3500 meters, as the sensitivity is only improved by a factor of $\sim1.2$.

With a volume of 10 $\km^3$, deployed in the deep South China Sea close to the equator, NEON has the capability of detecting high-energy neutrinos, with large effective area, a wide energy range from TeV to PeV, and full sky coverage.
The angular resolution has been greatly improved, reaching 0.1 degree at 100 TeV, further reducing the background event numbers and improving the sensitivity. These advantages will provide unprecedented precision for unraveling the origins of neutrinos.

Lately IceCube has confirmed our Galaxy is a guaranteed source of high-energy neutrinos, and LHAASO has revealed about 90 very high energy gamma-ray sources. Galactic sources are one of our main science objectives. These sources are usually characterized by spatial extension and high-energy cutoff. For the Galactic ridge, which is an extended neutrino source of around 2 degrees, the 5$\sigma$ discovery potential with a spectral index of 2.4 \cite{ANTARES:2022izu} can be achieved for an observation time of 5 years only.

Most of the high-energy astrophysical neutrino observations originate from extragalactic sources, some of which might be gamma-ray obscured. To make identification, time-integrated and time-dependent point source searches will be performed. Taking advantage of the high statistics and better angular resolution of NEON, we will have a great opportunity to detect these steady or transient neutrino sources, such as AGNs and TDEs. The detection of a doublet of neutrinos is definitely statistically significant. Assume a transient neutrino source with a flux spectrum with dependence on $E^{-2}$, and a temporal window of 500 seconds, so that the atomspheric neutrino background can be ignored. We estimate the \deleted{5$\sigma$} discovery potential to a doublet from this source, which is around $1\times10^{-7}$GeV cm$^{-2}$ s$^{-1}$. Once these signals are caught, the neutrino source will be confirmed, regardless of whether with or without other electromagnetic signals.

In summary, compared to the \replaced{currently constructed }{current} neutrino detector\added{, IceCube}, the effective area of NEON has been improved by up to one order of magnitude, as shown in the Figure \ref{fig:effareas} . NEON will largely increase the quality and quantity of neutrino observations, and explore the universe with a wider field of view. NEON would like to contribute to the global neutrino network and will have important synergies with the new generation of major astronomical and astroparticle observatories. 

\section*{Acknowledgements}
This work is supported by the National Natural Science Foundation of China (NSFC) grants 12261141691, 12205388, and 12005313. We thank the support from Fundamental Research Funds for the Central Universities, Sun Yat-sen University, No. 24qnpy123. 
\newpage
\bibliography{NEON_performance}
\bibliographystyle{unsrt}
\end{document}